\documentclass[pre,twocolumn,showpacs]{revtex4}
\usepackage{graphicx}
\newcommand{\beq}{\begin{equation}}
\newcommand{\eeq}{\end{equation}}
\newcommand{\beqa}{\begin{eqnarray}}
\newcommand{\eeqa}{\end{eqnarray}}
\newcommand{\ba}{\begin{array}}
\newcommand{\ea}{\end{array}} 
\begin{document} 

\title{Self-bound droplet of Bose and Fermi atoms in one dimension: \\
Collective properties in mean-field and Tonks-Girardeau regimes} 
\author{Luca Salasnich$^{1}$\footnote{Electronic address: 
salasnich@pd.infn.it; URL: http://www.padova.infm.it/salasnich}, 
Sadhan K. 
Adhikari$^{2}$\footnote{Electronic address: adhikari@ift.unesp.br; 
URL: http://www.ift.unesp.br/users/adhikari}, 
and Flavio Toigo$^{3}$\footnote{Electronic address: 	
toigo@pd.infn.it} }
\affiliation{$^1$CNISM and CNR-INFM, Unit\`a di Padova, 
Via Marzolo 8, 35131 Padova, Italy 
\\ 
$^{2}$Instituto de F\'isica Te\'orica, UNESP $-$ S\~ao Paulo 
State University, 
01.405-900 Sao Paulo, Sao Paulo, Brazil  
\\ 
$^{3}$Dipartimento di Fisica ``G. Galilei'' and CNISM, 
Universit\`a di Padova, Via Marzolo 8, 35131 Padova, Italy} 

\begin{abstract} 
We investigate a dilute mixture of 
bosons and spin-polarized fermions in one-dimension. 
With an attractive Bose-Fermi scattering length 
the ground-state is a self-bound droplet, i.e. 
a Bose-Fermi bright soliton where the Bose and Fermi clouds 
are superimposed. We find that the quantum fluctuations
stabilize the Bose-Fermi soliton such that the one-dimensional 
bright soliton exists for any finite attractive Bose-Fermi 
scattering length. We study density profile and collective excitations 
of the atomic bright soliton showing that they depend on the bosonic 
regime involved: mean-field or Tonks-Girardeau. 
\end{abstract}


\maketitle
 
\section{Introduction} 

Ultracold vapors of alkali-metal atoms, like 
$^{87}$Rb, $^{85}$Rb, $^{40}$K, $^{23}$Na, $^6$Li, $^7$Li, etc. 
are now actively studied in the regime 
of deep Bose and Fermi degeneracy 
\cite{stringari-book,ohara,greiner,jochim,bourdel}. 
Trapped Bose-Fermi mixtures, 
with Fermi atoms in a single hyperfine state, 
have been investigated by various authors both 
theoretically \cite{molmer,nygaard,viverit1,viverit2,das1,liu,wang} 
and experimentally 
\cite{roati,modugno,ospelkaus,bosfer,bosfer1,bosfer2}. 
Recently, it has been predicted that 
self-bound droplets, also called atomic bright solitons, 
can be formed within a mixture of degenerate Bose-Fermi 
gases provided the gases attract each other 
strongly enough and that there is an external transverse 
confinement \cite{karp,adh,konotop}. 
Formation of bright solitons in a dilute spin-polarized 
Fermi gas is prevented by Pauli repulsion. 
The formation of bright 
soliton in a Bose-Fermi mixture is related to the fact 
that the system can lower 
its energy by forming high-density regions (bright solitons) when the 
Bose-Fermi attraction is sufficient to overcome the Pauli repulsion
among Fermi atoms and any possible repulsion among the Bose atoms. 
A common point of these papers \cite{karp,adh,konotop}
is that the Fermi cloud is three-dimensional (3D). 
In fact, for not too strong Bose-Bose repulsion 
the transverse width of the Fermi component 
significantly exceeds the transverse width of 
the Bose component \cite{karp,adh,konotop}. 

In the strict one-dimensional (1D) regime, the Bose-Fermi mixture 
requires an appropriate theoretical description. 
The exponent of the power-law which describes the bulk energy 
of a Fermi gas as a function of its density 
depends on the dimensionality (see for instance \cite{sala-jmp}).
In addition, even at zero temperature, the 1D Bose gas can 
never be a true Bose-Einstein condensate 
due to phase fluctuations \cite{stringari-book,solitirussi}. 
For a repulsive 1D Bose gas, one must distinguish 
two regimes: a quasi-Bose-Einstein condensate (BEC) regime, 
well described by the 1D Gross-Pitaevskii equation 
with positive nonlinearity \cite{solitirussi}, 
and a Tonks-Girardeau (TG) regime at very low densities, 
where the 1D bosons behave as 1D ideal fermions \cite{lieb,girardeau0}. 
An attractive 1D Bose gas is instead well described by the Hartree 
mean-field theory \cite{mcguire,calogero}, i.e. 
the 1D Gross-Pitaevskii equation with 
negative nonlinearity \cite{sala-bmg}. 

The existence of the TG regime above has been recently 
experimentally confirmed \cite{TG1} in a study of the 1D 
degenerate $^{87}$Rb system. 
In a subsequent study of this system \cite{TG2}, the 1D Bose gas in 
the TG regime has been found to possess the peculiar 
property of not attaining a thermal equilibrium 
even after thousands of 
collisions. This is often termed fermionization of 1D bosons
in the TG regime. 
It is well-known that due to Pauli principle the 
spin-polarized trapped fermions do not interact at low temperature and 
hence fail to reach a thermal equilibrium necessary for evaporative cooling 
leading to a degenerate state. The necessary thermal equilibrium 
was attained only in  Bose-Fermi \cite{bosfer,bosfer1,bosfer2} or 
Fermi-Fermi \cite{ferfer} mixtures through 
collision between bosons and fermions or between fermions in different 
quantum states, respectively. 

In this paper we consider a Bose-Fermi mixture strongly confined 
by a 2D harmonic potential in the transverse cylindric 
radial coordinate. The ensuing effective 1D system is 
described in the quantum hydrodynamical approximation, 
i.e. the time-dependent density functional approach
based on real hydrodynamic variables or complex scalar fields. 
Quantum hydrodynamics is very useful for the study of static and collective 
properties of a Bose-Fermi mixture and it has been used successfully
in 3D \cite{adh,konotop,ska} for a description of bright and dark
solitons and collapse. 
We investigate the 1D mixture of bosons and spin-polarized fermions 
by using an effective 1D Lagrangian 
\cite{stringari-book,kolo,minguzzi,sala-lls}. 
A Gaussian variational approach is adopted to derive 
axial static and dynamical properties of the mixture with 
attractive Bose-Fermi scattering length ($a_{bf}<0$).
The solution of the variational scheme was found to be in satisfactory  
agreement with the accurate numerical solution of the hydrodynamic 
equations. We find that a self-bound droplet, i.e. a Bose-Fermi 
bright soliton, exists also for very small values 
of $|a_{bf}|$. In this case the axial width of the Fermi component 
is very large while the axial width of the Bose component 
depends on the sign and magnitude of the Bose-Bose 
scattering length $a_b$. Remarkably, the TG regime 
is essential to preserve a localized Bose-Fermi soliton 
for very small $|a_{bf}|$; in fact, for a repulsive Bose-Bose 
interaction in the quasi-BEC 
1D regime the theory predicts a minimum value of $|a_{bf}|$ 
below which the mixture is uniform, i.e. fully delocalized. 
For large values of $|a_{bf}|$ the Bose-Fermi system 
is self confined in a very narrow region 
and therefore the local axial densities of bosons and fermions strongly 
increase. We must remember, however, that 
above a critical axial density the Fermi system is 
no more strictly one-dimensional, and the same happens for 
repulsive bosons. 

The paper is organized as follows. 
In Sec. II we present the model used to study 
the degenerate Bose-Fermi system. Then we derive a set of coupled equations 
for the mixture starting from a Lagrangian density. In Sec. III, 
by using a Gaussian variational ansatz, we demonstrate 
that for an attractive Bose-Fermi interaction, 
the ground state of the model is a self-bound bright soliton 
(in the absence of an external longitudinal trap). 
In Sec. IV we present a study of the system in the quasi-BEC regime and 
in the TG regime for attractive 
Bose-Fermi interaction. These results are further explored 
in Sec. V considering a single Fermi atom inside the Bose cloud. 
In Sec. VI we consider the problem of coupled breathing oscillations of 
the Bose-Fermi system and calculate the frequencies of these oscillations. 
Finally, in Sec. VII we present a brief summary of our investigation 
and discuss the experimental conditions necessary 
to achieve the one-dimensional Bose-Fermi soliton. 

\section{Bose-Fermi Lagrangian for 1D hydrodynamics} 

We consider a mixture of $N_b$ 
bosons of mass $m_b$ and $N_f$ spin-polarized 
fermions of mass $m_f$ at zero temperature 
trapped by a tight cylindrically symmetric harmonic 
potential of frequency $\omega_{\bot}$ in the 
transverse direction. We assume factorization
of the transverse degrees of freedom.
It is justified in 1D confinement where, regardless
of the longitudinal behavior
or statistics, the transverse spatial profile is that
of the single-particle ground-state
\cite{das1,sala-lls,sala-reduce,sala-npse,das2}. 
The transverse width of the atom distribution  
is given by the characteristic harmonic length of the
single-particle ground-state: 
$a_{\bot j}=\sqrt{\hbar/(2m_j\omega_{\bot})}$, with $j=b,f$. 
The atoms have an effective 1D 
behavior at zero temperature if their 
chemical potentials are much smaller than 
the transverse energy $\hbar\omega_{\bot}$ \cite{das1,sala-lls}. 
\par 
We use a hydrodynamic effective Lagrangian 
to study the static and collective properties 
of the 1D Bose-Fermi mixture. 
In the rest of the paper all quantities are dimensionless.
In particular, lengths are in units of $a_{\bot b}$, linear densities  
in units of $a_{\bot b}^{-1}$, times in units of $\omega_{\bot}^{-1}$ 
and energies in units of $\hbar\omega_{\bot}$. 
The Lagrangian density ${\cal L}$ of the mixture reads 
\beq  
\label{z1} 
{\cal L} = {\cal L}_b + {\cal L}_f + {\cal L}_{bf} \; . 
\eeq 
The term ${\cal L}_b$ is the bosonic Lagrangian, defined as 
\beq
\label{z2}
{\cal L}_b  = \psi_b^* \big( i\partial_t + \partial_z^2\big) \psi_b   
- |\psi_b|^6 G\left({g_b \over 2|\psi_b|^2}\right)
- V_b |\psi_b|^2 \; , 
\eeq
where $\psi_b(z,t)$ is the hydrodynamic field of the Bose gas, 
such that $n_b(z,t)=|\psi_b(z,t)|^2$ is the 1D density 
and $v_b(z,t)=i\partial_z \ln{\left( \psi_b(z,t)/|\psi_b(z,t)| \right)}$ 
is its velocity. 
Here $g_b=2a_b/a_{\bot b}$ is the scaled inter-atomic 
strength with $a_b$ the Bose-Bose scattering length. 
We take $|g_b|<1$ to avoid the 
confinement-induced resonance \cite{olshanii}. 
Interacting bosons are one-dimensional if 
$g_b\, n_b\ll 1$ \cite{sala-lls,sala-reduce,sala-npse}. 
For $x>0$ the function $G(x)$ 
is the so-called Lieb-Liniger function, 
defined as the solution of a Fredholm equation 
and such that $G(x)\simeq x$ for $0< x \ll 1$ 
and $G(x)\simeq \pi^2/3$ for $x\gg 1$ \cite{lieb}. 
For $x<0$ we set $G(x)=x$ \cite{sala-bmg,sala-lls}. 
$V_b(z)$ is the longitudinal external potential 
acting on the bosons. In the static case the  
Lagrangian density ${\cal L}_b$ reduces exactly 
to the energy functional recently introduced by 
Lieb, Seiringer and Yngvason \cite{lieb2}. 
In addition, ${\cal L}_b$ has been successfully 
used to determine the collective oscillation 
of the 1D Bose gas with longitudinal 
harmonic confinement \cite{sala-lls}. 
\par
The fermionic Lagrangian density ${\cal L}_f$ is  given instead by 
\beq
{\cal L}_f = \psi_f^* \big( \partial_t 
+ \lambda_m \partial_z^2\big) \psi_f 
- {\pi^2 \lambda_m \over 3} |\psi_f|^6
- V_f |\psi_f|^2  \, , 
\eeq
where $\lambda_m = m_b/m_f$ and 
$\psi_f(z,t)$ is the hydrodynamic field 
of the 1D spin-polarized 
Fermi gas, such that $n_f(z,t)=|\psi_f(z,t)|^2$ 
is the 1D fermionic density and 
$v_f(z,t)=i \lambda_m \partial_z 
\ln{\left( \psi_f(z,t)/|\psi_f(z,t)| \right)}$ 
is the velocity of the Fermi gas. 
The non-interacting fermions are one-dimensional if
$(\pi^2\lambda_m/2) n_f^2 \ll 1$ \cite{das1}.  
$V_f(z)$ is the longitudinal external potential acting 
on fermions. In the static case and with $V_f(z)=0$ 
the Lagrangian ${\cal L}_f$ gives the correct energy 
density of a uniform and non-interacting 1D Fermi gas. 
More generally, the Euler-Lagrange equation of 
${\cal L}_f$ yields the hydrodynamic equations 
of the 1D Fermi gas \cite{minguzzi}. 
\par 
Finally, the Lagrangian density ${\cal L}_{bf}$ 
of the Bose-Fermi interaction reads 
\beq \label{z3}  
{\cal L}_{bf} = - g_{bf} \, 
|\psi_b|^2 |\psi_f|^2 \; , 
\eeq 
where $g_{bf}=2 a_{bf}/a_{\bot b}$ is the scaled inter-atomic 
strength between bosons and fermions, with $a_{bf}$ 
the Bose-Fermi scattering length \cite{das1}. 
\par  
Euler-Lagrange equations of the Lagrangian 
${\cal L}$ provide the two coupled partial 
differential equations for $\psi_b$ and $\psi_f$: 
\begin{eqnarray}
i \partial_t \psi_b &=& \biggr[ - \partial_z^2 
+ 3|\psi_b|^4 G\biggr({g_b\over 2|\psi_b|^2}\biggr) \nonumber \\
&-& {1\over 2}g_b|\psi_b|^2 
G'\biggr({g_b\over 2|\psi_b|^2}\biggr) 
+ V_b + g_{bf} |\psi_f|^2 \biggr] \psi_b  \; , 
\label{str1}
\end{eqnarray} 
\begin{eqnarray}
i \partial_t \psi_f = 
\left[ - {\lambda_m}\partial_z^2 + 
\pi^2 \lambda_m |\psi_f|^4 + V_f 
+ g_{bf} |\psi_b|^2 \right] \psi_f  \; . 
\label{str2} 
\end{eqnarray}
For $g_{bf}=0$ and $0< g_b < 1$, 
the first partial differential equation (\ref{str1})  
reduces, in the regime $g_b/n_b \ll 1$, 
to the familiar mean-field 1D Gross-Pitaevskii equation 
\cite{stringari-book}, i.e. to the 1D cubic nonlinear 
Schr\"odinger equation describing a quasi-BEC. 
Instead, in the regime where everywhere $g_b/n_b \gg 1$, 
Eq. (\ref{str1}) for bosons becomes the 
quintic nonlinear Schr\"odinger equation
proposed by Kolomeisky {\it et al.} \cite{kolo} 
for the dynamics of a TG gas, which is formally 
equivalent to Eq. (\ref{str2}) describing the 1D noninteracting Fermi gas. 
Actually, Girardeau and Wright \cite{girardeau} 
have shown that this quintic 
nonlinear Schr\"odinger equation overestimates the 
coherence in interference patterns at small number 
of particles. Nevertheless, Minguzzi {\it et al.} 
\cite{minguzzi} have found that this quintic 
equation is quite accurate 
in describing the density profile and the 
collective oscillations of the 1D ideal Fermi gas 
with longitudinal harmonic confinement. 
If we define $G(x)=x$ for $x<0$ then, when $g_{bf}=0$ and $g_b<0$, 
the Eq. (\ref{str1}) reduces to the mean-field 
1D Gross-Pitaevskii equation with attractive (negative) 
nonlinearity, which describes quite 
accurately the attractive 1D Bose gas \cite{calogero,sala-bmg}. 

\section{Self-bound solution: Bose-Fermi bright soliton}

In the remaining part of the paper we set $V_b(z)=V_f(z)=0$
and investigate the case of a negative Bose-Fermi
scattering length ($g_{bf}<0$). We use a time-dependent 
variational ansatz for the fields $\psi_j(z,t)$ 
to determine the conditions under which a self-bound droplet 
of 1D bosons and fermions exists. 
In particular we investigate the two main regimes 
of 1D bosons: the quasi-BEC regime and the TG regime. 
For the two fields $\psi_j(z,t)$, with $j=b,f$, 
we use the following Gaussian ansatz  
\beq 
\label{z4} 
\psi_j = {N_j^{1/2} \over \pi^{1/4} \sigma_j^{1/2}} 
\exp{\left( -{(z-z_j)^2\over 2 \sigma_j^2} 
+ i\phi_j z + i \theta_j z^2 \right)} \; , 
\eeq
where the time-dependent variational 
parameters are the longitudinal widths $\sigma_j(t)$, 
the centers of mass $z_j(t)$, and the slopes $\phi_j(t)$ 
and curvatures $\theta_j(t)$ of the phase. 
It is obvious that the tails of the Gaussian 
$n_b(z,t)=|\psi_b(z,t)|^2$ given by Eq. (\ref{z4})
are locally in the TG regime but, 
in our terminology, a non-uniform cloud of bosons is 
in the TG regime only if {\it everywhere} 
its local density $n_b(z,t)$ satisfies the condition
$g_b/n_b(z,t)\gg 1$. 

We insert the Gaussian fields $\psi_j(z,t)$ into the Lagrangian 
${\cal L}$ and integrate over the 
spatial variable $z$ and get an effective 
Lagrangian \cite{perez,sala0} which depends 
on $\sigma_j(t)$, $z_j(t)$, $\phi_j(t)$, $\theta_j(t)$ 
and their time derivatives. 
By writing the eight  Euler-Lagrange equations 
one finds that the slopes $\phi_j(t)$ and the 
curvatures $\theta_j(t)$ of the fields $\psi_j(z,t)$ 
can be obtained from the widths $\sigma_j(t)$ 
and the center coordinates $z_j(t)$ through the equations 
\beq 
\phi_j= -{\dot z_j} - 2 \theta_j z_j \; , \quad 
\theta_j = - {{\dot \sigma_j}\over 2\sigma_j} \; , 
\eeq 
with $j=b,f$. The equations of motion of the 
parameters $\sigma_j(t)$ and $z_j(t)$ do not depend 
on the phase parameters $\phi_j(t)$ and $\theta_j(t)$ \cite{perez,sala0}.  
They are the ``classical'' equations of motion of a system 
with effective Lagrangian 
\beq  
L= T - E \; ,  
\label{eff-lagrangian}
\eeq 
where 
\beq 
T = {N_b \over 2 } 
\left( {\dot \sigma_b}^2 + 2 {\dot z_b}^2 \right) + 
{N_f \lambda_m \over 2 } 
\left( {\dot \sigma_f}^2 + 2 {\dot z_f}^2 \right) 
\label{eff-kinetic}
\eeq 
is the effective kinetic energy and 
\beq 
\label{e1}
E = E_b + E_f + E_{bf}, 
\eeq 
is the effective potential energy of the system. 
The term $E_b$ involves a complicated integral 
of the Lieb-Liniger function $G(x)$, namely 
\beq 
\label{e2}
E_b = {N_b \over 2 \sigma_b^2} + 
{N_b^3 \over \pi^{3/2} \sigma_b^2} 
\int_{-\infty}^{+\infty} e^{-3y^2} 
G\biggr({g_b \sigma_b \over 2 N_b}\, e^{y^2} \biggr) \, dy 
\; .   
\eeq
The other two terms, $E_f$ and $E_{bf}$, are given by 
\beq 
\label{e3}
E_f = {N_f \lambda_m \over {2\sigma_f^2}} + 
{N_f^3 \pi \lambda_m \over 3    \sqrt{3}
\sigma_f^2 }  
 \; , 
\eeq
and 
\beq 
\label{e4}
E_{bf}= {g_{bf}N_b N_f\over \sqrt{\pi} 
\sigma_{bf} } \exp{\Big(-{(z_b-z_f)^2\over 
\sqrt{\sigma_b^2 + \sigma_f^2}} \Big)} \; . 
\eeq  
We stress that the potential energy (\ref{e1}) of the 
effective Lagrangian (\ref{eff-lagrangian}) can be 
easily obtained from  ansatz (\ref{z4}) without 
including the phase parameters $\phi_j(t)$ and $\theta_j(t)$. 
On the contrary, to get the kinetic energy term (\ref{eff-kinetic}) 
it is necessary to include in the ansatz the four phase 
parameters of Eq. (\ref{z4}) \cite{perez,sala0}. 
The kinetic term is essential to calculate the dynamical 
properties of the mixture, such as the collective oscillations considered 
in Sec. VI. 

\begin{figure}[tbp]
\includegraphics[width=.9\linewidth]{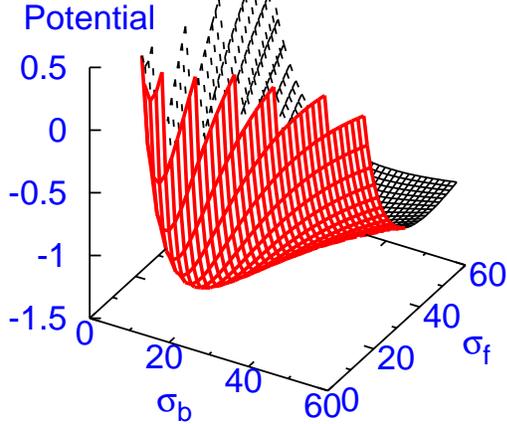}
%\centerline{\epsfig{file=solitonBF-f1.eps,width=8.cm,clip=}}
\caption{(Color online).
Effective potential energy $E$ of Eq. (\ref{e1}) 
as a function of soliton widths $\sigma_b$
and $\sigma_f$ for parameters:
$\lambda_m=1, N_b=100,N_f=10,g_b=0.01$ and $g_{bf}=-0.2$. 
The potential has a minimum at $\sigma_b=16.90$ and $\sigma_f=21.84$. 
Lengths are in units 
of $a_{\bot b}=\sqrt{\hbar/(2m_b\omega_{\bot})}$ and energy
in units of $\hbar \omega_{\bot}$.}
\end{figure}

The stable stationary state of the system is found 
by minimizing the effective potential energy:  
\beq 
\label{e51}
{\partial E\over \partial z_j} =0 \; , \quad j=b,f \; ,
\eeq 
\beq \label{e52}
{\partial E\over \partial \sigma_j} =0 \; , \quad j=b,f \; .
\eeq 
Equations (\ref{e51}) lead to $z_b=z_f$ and without 
loss of generality we set $z_b=z_f=0$. 
Equations (\ref{e52}) can then be rewritten as
$$
1 + {2 N_b^2 \over \pi^{3/2}} 
\int_{-\infty}^{+\infty} e^{-3y^2} \, 
G\biggr({g_b \sigma_b \over 2 N_b}\, e^{y^2} 
\biggr) \, dy 
$$
\beq 
-{g_b N_b \, \sigma_b \over 2\pi^{3/2}} 
\int_{-\infty}^{+\infty} e^{-2y^2} \, 
G'\biggr( {g_b \sigma_b \over 2 N_b} \, e^{y^2} \biggr) 
= {-g_{bf} N_f \, \sigma_b^4 
\over \sqrt{\pi} \sigma_{bf}^3}  \; , 
\label{pp1} 
\eeq 
\beq 
\lambda_m + 
{2 N_f^2 \pi \lambda_m \over 3 \sqrt{3}} = 
{-g_{bf} N_b \, \sigma_f^4 \over \sqrt{\pi} 
%\left( \sigma_b^2 + \sigma_f^2 \right)^{3/2} 
\sigma_{bf}^3
} \; , 
\label{pp2} 
\eeq
where $\sigma_{bf}=\sqrt{\sigma_b^2 + \sigma_f^2}$. 
From Eqs. (\ref{pp1}) and (\ref{pp2}) one can determine the widths 
$\sigma_b$ and $\sigma_f$ at  equilibrium. 
Stability requires that the Hessian matrix 
of the second partial derivatives of the effective potential energy 
$E(\sigma_j)$ is positive definite; equivalently the Gaussian curvature 
\beq 
\label{k1}
K_G = {\partial^2 E\over \partial \sigma_b^2} 
{\partial^2 E\over \partial \sigma_f^2} - 
\left( {\partial^2 E\over \partial \sigma_b\partial\sigma_f} \right)^2 
\eeq  
must be positive. An inspection of Eqs. (\ref{pp1}) 
and (\ref{pp2}) shows that there are stable solutions 
only for $g_{bf}\leq 0$. For $g_{bf}=0$ and $g_b>0$, the solutions 
are $\sigma_b=\sigma_f=+\infty$, corresponding to infinitely 
extended, uniform bosonic and fermionic clouds, while for $g_{bf}=0$ 
and $g_b<0$, the fermionic 1D density is uniform 
while the bosonic cloud is localized with 
$\sigma_b=2\sqrt{2\pi}/(|g_b|N_b)$. For $g_{bf}<0$ 
the Eqs. (\ref{pp1}) and (\ref{pp2}) must be solved numerically. 
In the numerical calculations the Lieb-Liniger function $G(x)$ is modeled 
by an efficient Pad\`e approximant based on the exact numerical 
determination of $G(x)$ \cite{pade}. 

As the effective potential of the problem is $E$ of Eq. (\ref{e1}), the 
equation (\ref{e52}) together with condition 
implicit in Eq. (\ref{k1}) minimizes the effective potential as a 
function of the two widths $\sigma_b$ and $\sigma_f$. A 
typical plot of 
the potential for $N_b=100$, $N_f=10$, $g_b=0.01$,  $g_{bf}=-0.2$, 
and $\lambda_m=1$ is shown in Fig. 1. Stable oscillations of the system 
are possible around the minimum. 
We shall study different features of these 
oscillations  in the following.
 
As previously stressed, bosons are one-dimensional 
under the condition $g_b n_b \ll 1$, which corresponds to 
$\sigma_b\gg g_b N_b/\sqrt{\pi}$. For a quite large width, 
namely for $\sigma_b \gg N_b/(\sqrt{\pi}g_b)$, the bosons 
enter in the TG regime, where $g_b/n_b \gg 1$. 
The fermions are instead one-dimensional under the condition 
$(\pi^2\lambda_m/2)n_f^2\ll 1$, which corresponds to 
$\sigma_f \gg N_f \sqrt{\lambda_m\pi/2}$. 

\begin{figure}[tbp]
\includegraphics[width=.9\linewidth]{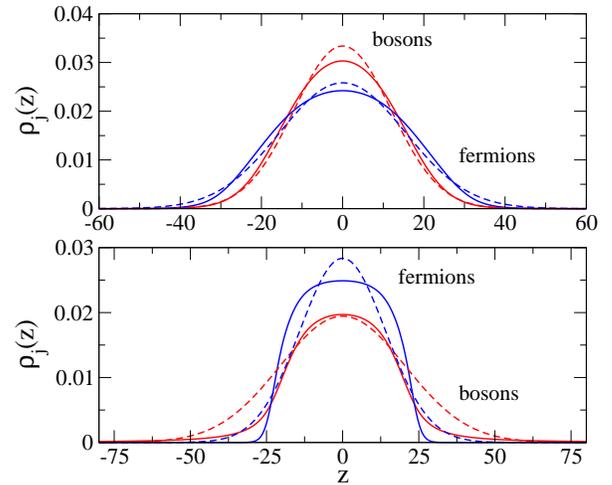}
%\centerline{\epsfig{file=solitonBF-f2.eps,width=8.cm,clip=}}
\caption{(Color online). 
Probability density $\rho_j(z)=|\psi_j(z)|^2/N_j$ of 
bosons and fermions in the self-bound mixture 
with $g_b=0.01$ and $g_{bf}=-0.2$. 
In the upper panel: $N_b=100$ and $N_f=20$; 
in the lower panel: $N_b=300$ and $N_f=10$. 
Solid lines: numerical results. Dashed lines: variational results. 
Units as in Fig. 1.} 
\end{figure}

We shall base the present study on the Gaussian variational approach 
described above, which, like any variational approach, should be 
reliable for the Bose-Fermi ground state studied in this paper. 
Moreover, the analytical variational solution provides interesting
physical insight into the problem, as we shall see in the following. 
Also, the actual numerical solution of the full coupled dynamics is 
pretty complicated to implement for all cases reported in this paper 
in the various parameter ranges. 
Nevertheless, we find it worthwhile to compare 
the solution of the variational scheme with the accurate 
numerical solution of Eqs. (\ref{str1}) and (\ref{str2}) in certain 
cases. We solved these numerically, 
using a imaginary-time integration method based on the 
finite-difference Crank-Nicholson scheme, as described 
in Ref. \cite{sala-metodi}. We discretize the mean-field equations  
using a time step $\Delta t=0.05$ and a space step $\Delta z =0.05$, 
and $z\in [-L/2,L/2]$ with $L=2000$. The boundary conditions 
are $\psi_j(-L/2)=\psi_j(L/2)=0$, with $j=b,f$. 
We start with broad Gaussians as initial wave functions. 
In the course of the imaginary-time evolution the self-bound 
mixture is quickly formed but, due to strong Pauli repulsion among
identical spin-polarized fermions, 
the fermionic density profile extends to many 
hundredths of length's units. It is then essential to take a very 
large space interval $[-L/2,L/2]$ of integration 
to see that these long tails of the fermionic cloud are indeed 
decaying to zero. 

In Fig. 2 we plot two sets of numerical results for the 
probability density in the quasi-BEC 1D regime. 
The figure shows that, for fixed values of the interaction 
strengths $g_b$ and $g_{bf}$, the axial width of the 
bosonic probability density becomes larger than the 
fermionic one by reducing the number 
$N_f$ of fermions and increasing the number $N_b$ of bosons. 
The figure shows that the variational approach 
can be used to give a reasonable estimation of the axial 
widths of the two clouds. 

We now turn to a study of the 
Bose-Fermi system by using the variational approach which 
enables us to explore quite easily all regimes 
and extract physically interesting analytical results.  
\begin{figure}[tbp]
\includegraphics[width=.9\linewidth]{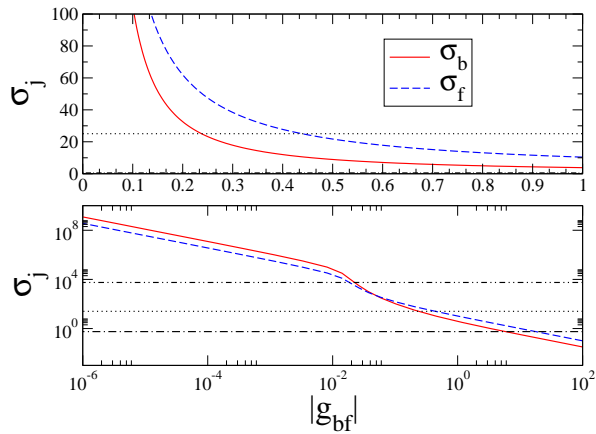}
%\centerline{\epsfig{file=solitonBF-f3.eps,width=8.cm,clip=}}
\caption{(Color online). 1D Self-bound Bose-Fermi droplet with repulsive 
bosons ($g_b>0$). Axial widths $\sigma_b$
and $\sigma_f$ of the bosonic and fermionic widths
as a function of the Bose-Fermi interaction strength $|g_{bf}|$,
with $g_{bf}<0$. Mixture parameters: $N_B=100$, $N_F=20$, 
$\lambda_m=1$ and $g_b=0.01$. Upper panel: linear-linear scale. 
Lower panel: log-log scale. Between the dot-dashed line and 
the dot-dot-dashed line the bosons are in the quasi-BEC 1D regime, 
while above the dot-dot-dashed line they are in the 
TG regime. Fermions are in the 1D regime above the 
dotted line. Units as in Fig. 1.}
\end{figure}
With the intention of illustrating the TG 
regime and the quasi-BEC Gross-Pitaevskii 
regime for bosons and 1D regime for fermions 
for a specific set of Bose-Fermi parameters. 
In Fig. 3 we plot  $\sigma_b$ (solid line) 
and   $\sigma_f$ (dashed line)  from Eqs. (\ref{pp1}) and (\ref{pp2})
as a function of $|g_{bf}|$, with $g_{bf}<0$ 
for the values $N_b=100$, $N_f=20$, 
$g_b=0.01$, and $\lambda_m=1$ of the parameters. 
In the upper panel  the linear-linear scale is employed while in the
lower panel we report the same results on a log-log scale, to better visualize 
the TG regime and the quasi-BEC regimes. 
When their width takes values between 
$\sigma_b=g_bN_b/\sqrt \pi\simeq 0.56$ (dot-dashed line) 
and  $\sigma_b=N_b/(g_b\sqrt \pi) \simeq 5600$ ( dot-dot-dashed line)  
bosons are in the quasi-BEC 1D regime, 
while   they are in the
TG 1D regime if $\sigma_b>5600$ 
(above the dot-dot-dashed line). 
Fermions are in the 1D regime when 
$\sigma_f>N_f\sqrt{\lambda_m\pi/2}\simeq 25$, i.e. above the dotted line. 
For large $|g_{bf}|$, both the widths tend to zero corresponding to 
narrow soliton(s). 

\section{Quasi-BEC 1D regime and Tonks-Girardeau regime} 

The expression for the effective Lagrangian of the Bose-Fermi system 
given by Eqs. (\ref{e1}) $-$ (\ref{e4}) is greatly simplified 
in the quasi-BEC 1D regime and also 
in the TG regime. 
In the quasi-BEC 1D regime one has $G(x)\simeq x$ 
and the expression for the effective energy $E$ of Eq. (\ref{e1}) 
can be evaluated analytically and written as 
\beq \label{k2}
E = {1\over 2} {\alpha\over \sigma_b^2} 
+ {\beta \over \sigma_b} + 
{1\over 2} {\gamma \over \sigma_f^2} 
- {\delta \over \sqrt{\sigma_b^2 + \sigma_f^2} } \; , 
\label{simple-bec} 
\eeq 
where $\alpha=N_b$, $\beta =g_bN_b^2/(2\sqrt{2\pi})$, 
$\gamma =\lambda_m N_f [1+ 2\pi N_f^2/(3\sqrt{3})]$, and  
$\delta =|g_{bf}|N_bN_f/\sqrt{\pi}$ with $g_{bf}=-|g_{bf}|$. 
Then at equilibrium one finds from Eq. (\ref{e52}) 
that $\sigma_f$ and $\delta$ can be written 
as functions of $\sigma_b$:  
\beq 
\label{x1}
\sigma_f = \gamma^{1/4} {\sigma_b\over 
(\alpha + \beta \sigma_b)^{1/4} } 
\eeq 
and also 
\beq \label{x2}
\delta = ({\alpha \over \sigma_b} + \beta ) 
\left(1+\sqrt{\gamma \over \alpha + \beta \sigma_b} \right)^{3/2}  \; . 
\eeq
Eq. (\ref{x2}) implies that for any finite $\sigma_b$ and $\sigma_f$ 
one has $\delta > \beta$, i.e. 
$|g_{bf}|> |g_{bf}|_{min}= \sqrt{2}g_bN_b/(4 N_f)$. 
Thus, for $g_b>0$ the Bose-Fermi bright soliton exists only for 
$|g_{bf}|>|g_{bf}|_{min}$, while the system will not be bound 
($\sigma_b\to\infty$, $\sigma_f\to \infty$) 
for $|g_{bf}|< |g_{bf}|_{min}$. 
This last situation is however unphysical 
because for a very large $\sigma_b$ the system 
enters in the TG regime and Eq. (\ref{simple-bec}) 
is no more valid. 

When the Bose-Bose scattering length is zero ($g_b=0$) 
then $|g_{bf}|_{min}=0$. 
For $g_b<0$ (attractive Bose-Bose interaction), 
$\sigma_b$ is finite even for $|g_{bf}|_{min}=0$, 
but it cannot exceed the value 
$\sigma_b=2\sqrt{2\pi}/(|g_b|N_b)$ with the 
corresponding value of $\sigma_f$ infinitely large. 
When $|g_{bf}|=0$ the bosonic cloud is localized 
($\sigma_b=2\sqrt{2\pi}/(|g_b|N_b)$)
while the fermionic cloud is fully delocalized ($\sigma_f=+\infty$)
in the axial direction. 

Coming back to the case of a repulsive Bose-Bose scattering length ($g_b>0$), 
we observe that, after fixing $g_b$ and $N_b$, by reducing $|g_{bf}|$ 
the bosonic width $\sigma_b$ increases and 
the system enters in the TG regime, where $G(x)\simeq \pi^2/3$ 
and the effective energy $E$ of Eq. (\ref{e1}) can also be evaluated 
analytically as 
\beq \label{x3}
E = {1\over 2} {\alpha \over \sigma_b^2} + \frac{1}{2}
{{\tilde \beta}\over \sigma_b^2} + 
{1\over 2} {\gamma \over \sigma_f^2} - 
{\delta \over \sqrt{\sigma_b^2 + \sigma_f^2} } \; . 
\label{simple-tg} 
\eeq 
The quantities $\alpha$, $\gamma$ and $\delta$ are the same 
as in Eq. (\ref{simple-bec}) while 
${\tilde \beta}=2 N_b^3\pi /(3\sqrt{3})$. In this case 
the widths $\sigma_b$ and $\sigma_f$ at the equilibrium 
can also be determined analytically. They are 
\beq \label{x4}
\sigma_b = {\alpha +{\tilde \beta} \over \delta} 
\left(1+\sqrt{\gamma \over (\alpha +{\tilde \beta})} \right)^{3/2} 
\eeq
\beq \label{x5}
\sigma_f = \gamma^{1/4} 
{(\alpha +{\tilde \beta})^{3/4} \over \delta} 
\left( 1+\sqrt{\gamma \over 
(\alpha+{\tilde \beta })} \right)^{3/2} \; ,  
\eeq 
and both increase as  $\delta$ decreases, 
and only when $\delta=|g_{bf}|=0$ 
the two widths become infinitely large. Thus we conclude 
that for any finite value of $|g_{bf}|$ and $g_b>0$ there exists 
a Bose-Fermi bright soliton, the existence of this bright soliton 
being guaranteed by the behavior of the bosonic energy term 
in the TG regime. 

\begin{figure}[tbp]
\includegraphics[width=.9\linewidth]{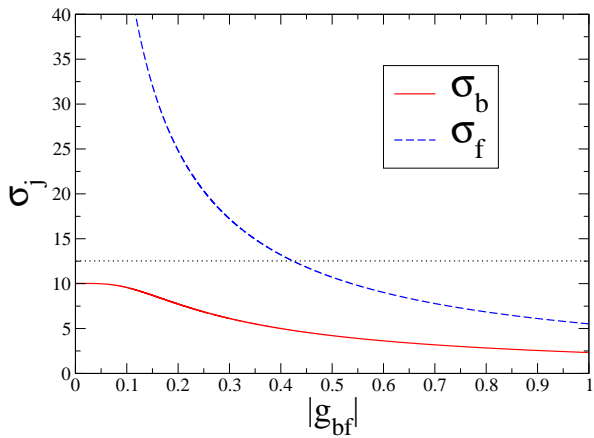}
%\centerline{\epsfig{file=solitonBF-f4.eps,width=8.cm,clip=}}
\caption{(Color online). 
1D Self-bound Bose-Fermi droplet with attractive bosons 
($g_b<0$). Axial widths $\sigma_b$ and $\sigma_f$ of the bosonic and 
fermionic widths as a function of the Bose-Fermi 
interaction strength $|g_{bf}|$,
with $g_{bf}<0$. Mixture parameters: $N_b=50$, $N_f=10$,
$\lambda_m=1$ and $g_b=-0.01$. 
Fermions are in the 1D regime above the dotted line. 
Units as in Fig. 1.}
\end{figure}

\par
In Fig. 4 we show the axial widths $\sigma_b$ and
$\sigma_f$ of the Bose-Fermi mixture
with an attractive Bose-Bose scattering length ($g_b<0$). 
We choose $N_b=50$, $N_f=10$, and $g_b=-0.01$ and plot
the widths as a function of $|g_{bf}|$. 
The figure shows that, as $|g_{bf}|$ goes to zero, 
the Fermi width $\sigma_f$ is much larger than the Bose 
width $\sigma_b$. In fact, as previously shown, 
at $|g_{bf}|=0$, one finds $\sigma_f=+\infty$ while 
$\sigma_b= 2\sqrt{2\pi}/(|g_b|N_b|)$. 
For large values of $|g_{bf}|,$ the Fermi width $\sigma_f$ quickly 
decreases and reaches the value $\sigma_f =
N_f \sqrt{\lambda_m \pi/2}\approx 12.5$ 
(dotted line) below which the Fermi 
system is no more strictly one-dimensional. 

\section{A single fermionic atom in the Bose cloud} 

The existence of the Bose-Fermi bright soliton 
for both attractive and repulsive Bose-Bose interaction, 
provided their is an attractive Bose-Fermi interaction, 
even vanishingly small, is due to the 1D effect of quantum 
fluctuations described by the Lieb-Liniger 
function $G(x)$. To understand this result further we consider the case 
of a single fermionic atom ($N_f=1$) interacting 
with the Bose cloud. In this case the exact stationary 
Schr\"odinger equation of the single-particle wave function 
$\psi_f(z)$ of the fermion is given by 
\beq \label{x6} 
\Big( - {\lambda_m}\partial_z^2 - |g_{bf}| |\psi_b|^2 \Big) 
\psi_f = \epsilon\, \psi_f \; , 
\eeq
where $\epsilon$ is the energy 
of the fermionic bound state 
under the effective potential well 
$V_{eff}(z)=-|g_{bf}||\psi_b(z)|^2$ determined 
by the Bose cloud. 
We know that for $g_b<0$ the pure 1D Bose system supports 
a bright soliton described by 
$\psi_b(z) = {(|g_b| N_b)^{1/2} \over {2} } 
\mbox{sech}\biggr( {|g_b| z\over {2} }\biggr)$.
Guided by this result, in the present case of repulsive bosons 
($g_b>0$) attracted by  a single fermion $g_{bf}<0$ we adopt
 for the bosonic field 
$\psi_b(z)$ the following ansatz 
\beq \label{x7} 
\psi_b(z) = {N_b^{1/2} \over (2 \xi )^{1/2} } 
\mbox{sech}\biggr( {z\over \xi }\biggr) \; , 
\eeq 
with $\xi$ the variational width of the bosonic cloud. 
This ansatz is also suggested by the 
existence of the analytical solution for the eigenvalue 
of the corresponding Eq. (\ref{x6}) for the fermionic bound state. 
One finds  \cite{landau}:   
\beq \label{y1} 
\epsilon(\xi ) = - {1\over 2\xi^2} 
\left( 1 + |g_{bf}|N_b \xi  - \sqrt{1+2|g_{bf}|N_b \xi} \right)^2 \; . 
\eeq 
The energy $\epsilon(\xi )$ is the sum of the kinetic energy 
of the fermionic atom and of the interaction energy between 
the fermionic atom and the bosonic cloud.

The total energy of the system is then the effective energy of the 
Bose cloud calculated from the Lagrangian density (\ref{z2})
added to the energy $\epsilon(\xi)$ of the Fermi atom and 
is given by 
\beq  \label{xx1}
E = {N_b\over 3\xi^2} + {N_b^3\over 8\xi^2}
\int_{-\infty}^{+\infty} \mbox{sech}^6(y) \, 
G\biggr({g_b\xi \over N_b \mbox{sech}^2(y)}\biggr) \, dy
+ \epsilon(\xi ) \; .
\eeq
The existence of the Bose-Fermi bright soliton
depends on the existence of a local minimum
of this total energy $E=E(\xi)$ of the system. 
From Eq. (\ref{xx1}) we find that $E(\xi) \sim - {|g_{bf}|N_b\over 2\xi}$
as $\xi\to +\infty$, since for large $\xi$'s
$G\big({g_b\xi \over N_b 
\mbox{sech}^2(y)}\big)\simeq {\pi^2\over 3}$ for any value of $y$.
Thus, for $\xi \to +\infty$, the energy
goes to zero through negative values. 
Again from Eq. (\ref{xx1}) we see that 
$E(\xi)\to +\infty$ as $\xi\to 0$  and therefore we conclude that 
the function $E(\xi)$, being positive at the origin 
and vanishingly small, but negative, at large $\xi$'s, 
must possess a negative local minimum
that is also the global minumum of the energy. This implies that there is
always a finite value of $\xi$ that minimizes the total energy.
As previously stressed, this behavior strictly
depends on the properties of the Lieb-Liniger function $G(x)$
for large $x$ and, as a consequence, the Bose-Fermi bright
soliton exists for any negative value $g_{bf}$ even for $N_f=1$.

\begin{figure}
\vskip 0.1cm
\includegraphics[width=.9\linewidth]{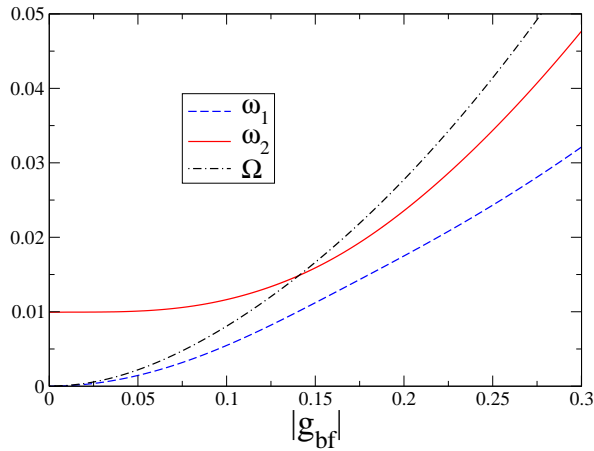}
%\centerline{\epsfig{file=solitonBF-f5.eps,width=8.cm,clip=}}
\caption{(Color online). 
Collective breathing frequencies $\omega_1$ and $\omega_2$ 
of the self-bound Bose-Fermi droplet with attractive bosons 
($g_b<0$). $\Omega$ is the harmonic frequency of 
the displacement $|z_1-z_2|$ of the centers of mass 
of the Bose and Fermi clouds. Parameters as in Fig. 4. 
Frequencies are in units of the frequency $\omega_{\bot}$ 
of transverse harmonic confinement.} 
\end{figure} 

\section{Collective oscillations of the Bose-Fermi bright soliton} 
 
After having established the existence of stationary bright solitons
in a degenerate Bose-Fermi mixture, we study two types of 
small oscillations of this system around the stable 
equilibrium position and calculate their frequencies. 
The first is the stable breathing 
oscillation of the system around its mean position 
and the second describes the 
stable small oscillations once the centers of the Fermi and  Bose 
clouds are slightly displaced with respect to each other. 
Given the effective Lagrangian (\ref{eff-lagrangian}), 
the problem of small oscillations is solved 
by expanding the kinetic energy (\ref{eff-kinetic}) and the 
potential energy (\ref{e1}) around the equilibrium solution  
up to quadratic terms. In this way we extract \cite{sala0} 
the frequencies $\omega$ of the collective breathing modes 
of the Bose and Fermi clouds from the eigenvalue equation 
\beq 
\left( \begin{array}{cc}
{\partial^2 E\over \partial \sigma_b^2} & 
{\partial^2 E\over \partial \sigma_b \sigma_f} \\
{\partial^2 E\over \partial \sigma_b \sigma_f} & 
{\partial^2 E\over \partial \sigma_f^2}  
\end{array} \right) 
- \omega^2 
\left( 
\begin{array}{cc}
N_b & 0  \\ 
0   & N_f \lambda_m  
\end{array} 
\right) 
= 0 \, 
\eeq
where the partial derivatives must be calculated at the
equilibrium $(\sigma_b,\sigma_b)$, where $\sigma_b$ and
$\sigma_f$ are the Bose and Fermi widths
obtained from Eqs. (\ref{pp1}) and (\ref{pp2}). 
The two frequencies $\omega_1$ and $\omega_2$ then read 
\beq  
\label{y3}
\omega_{1,2} = 
\sqrt{ N_b {\partial^2 E\over \partial \sigma_f^2} 
+ \lambda_f N_f {\partial^2 E\over \partial \sigma_b^2}  
\pm \sqrt{\Delta}  \over  2\lambda_m N_b N_f } \; , 
\eeq
where 
\beq \label{y4}
\Delta = \left( N_b {\partial^2 E\over \partial \sigma_f^2}
+ \lambda_m N_f {\partial^2 E\over \partial \sigma_b^2} \right)^2 
- 4 \lambda_m N_b N_f K_G 
\eeq 
and $K_G$ is the Gaussian curvature of Eq. (\ref{k1}). 
\par 
In addition, by taking into account the ``reduced mass'' 
$\lambda_m N_bN_f/(N_b+\lambda_m N_f)$, one can derive the frequency 
$\Omega$ of harmonic oscillation of the 
relative distance $|z_b-z_f|$ (displacement) 
of the two clouds from the equation 
\beq 
{\lambda_m N_bN_f \over N_b +\lambda_m N_f} \Omega^2 = 
{\partial^2 E\over \partial (|z_b-z_f|)^2} \; . 
\eeq
The result is 
\beq \label{y5}
\Omega = \sqrt{ 2|g_{bf}| ({N_b + \lambda_m N_f} ) 
\over \pi (\sigma_b^2+\sigma_f^2)^{3/4} } \; ,  
\eeq
where again $\sigma_b$ and $\sigma_f$ are the Bose and Fermi widths 
at equilibrium, where $z_b=z_f$.  
 
In Fig. 5 we plot the frequencies $\omega_1$ and $\omega_2$ of the 
coupled breathing modes of the Bose and Fermi clouds 
by choosing the same parameters of Fig. 4, namely $N_b=50$, 
$N_f=10$, $\lambda_m=1$ and $g_b=-0.01$. The frequencies are 
plotted as a function of the Bose-Fermi strength $|g_{bf}|$. 
For $|g_{bf}|=0$, the frequencies are decoupled: 
$\omega_1$ is the frequency of the fermionic 
axial breathing mode and $\omega_2$ is the frequency 
of the bosonic axial breathing mode. Without a Bose-Fermi 
interaction the Fermi cloud is delocalized ($\sigma_f=\infty$) and its 
breathing frequency is $\omega_1=0$, while the Bose cloud 
remains 
localized (due to the  negative Bose-Bose strength $g_b$) 
and its breathing frequency $\omega_2$ remains finite and 
is equal to $\omega_2=g_b^4 N_b^4/(16\pi^2)$. 
Fig. 5 shows that both the breathing frequency $\omega_1$ and  
the harmonic frequency $\Omega$ of the displacement $|z_1-z_2|$ 
start from zero and grow as $|g_{bf}|$ increases. 

\section{Conclusion}  

We have studied a degenerate 1D Bose-Fermi mixture by using 
the quantum hydrodynamics. We find that for 
attractive Bose-Fermi interaction ($g_{bf}<0$)
the ground state of the system is a self-bound Bose-Fermi droplet. 
The nonexistence of a threshold in the strength
of an attractive Bose-Fermi interaction for the formation 
of a Bose-Fermi bright soliton in one dimension 
is confirmed in the case of a single 
Fermi atom immersed in a degenerate Bose gas with repulsive 
Bose-Bose interaction. 
We also calculate the frequencies of stable 
oscillation of the Bose-Fermi bright soliton.
Such a Bose-Fermi bright soliton is similar to a 
recently studied Bose-Bose bright soliton bound through an attractive  
interspecies interaction \cite{vpg}. 

In view of the recent experimental studies of a degenerate 1D 
$^{87}$Rb gas \cite{TG1,TG2} and the successful identification 
of the quasi-BEC and TG regime in it and the 
observation of degenerate Bose-Fermi mixture in $^6$Li-$^7$Li 
\cite{bosfer,bosfer1}, 
$^{40}$K-$^{87}$Rb \cite{roati}, $^6$Li-$^{23}$Na  \cite{bosfer2} etc. 
by different groups, the experimental realization 
of a Bose-Fermi bright soliton seems 
possible with present technology. 
The most attractive procedure seems to make use of an experimentally 
observed Feshbach resonance \cite{fesh} in a Bose-Fermi mixture. 
The 1D Bose-Fermi mixture must be created in an axial harmonic 
trap and the Bose-Fermi interaction must be turned from repulsive 
to attractive by manipulating a 
background magnetic field. At the same time the axial harmonic trap on 
the system should be removed. Upon removal of the axial trap, the 
result is the formation of a single or a train of bright 
solitons as in the experiment with the 
degenerate Bose system of $^7$Li atoms \cite{bosfer1} or as in a 
numerical simulation in a degenerate Bose-Fermi mixture \cite{karp}. 
By choosing numbers of atoms and inter-atomic strengths 
as suggested in the present paper, one obtains a single Bose-Fermi 
bright soliton and can study its static and dynamical properties. 

\acknowledgments
This work is partially supported by the FAPESP and CNPq of Brazil.


\begin{thebibliography}{99}
\bibitem{pade} The Lieb-Liniger function $G(x)$ is such that 
for $x\ll 1$ one finds $G(x)\simeq x+Bx^{3/2}+\eta x^2$, 
where $B=4/(3\pi)$ and $\eta=0.0648$ (from numerics). 
Instead for $x\gg 1$ one has $G(x)\simeq (\pi^2/2)x^2/(x+2)^2$. 
The Pad\`e approximant we use is 
$G(x)=(x+Ax^3)/(1+Bx^{1/2}+Cx(x+2)^2)$, where $A=\pi^2(B^2-\eta )/12$ 
and $C=(B^2- \eta)/4$. For further details see \cite{sala-lls}.
\end{thebibliography}
\end{document}